\documentclass{article}
\usepackage{natbib}
\setcitestyle{round}

\usepackage[affil-it]{authblk}
\usepackage[english]{babel}

\usepackage[letterpaper,top=2cm,bottom=2cm,left=3cm,right=3cm,marginparwidth=1.75cm]{geometry}

\usepackage{amsthm}
\usepackage{graphicx} 
\usepackage{longtable}
\usepackage{booktabs,multirow,tabularx} 
\usepackage{multicol} 
\usepackage{float}
\usepackage{amsmath, amssymb, mathtools}
\usepackage{supertabular,booktabs}
\usepackage{makecell}
\usepackage{caption}
\usepackage{setspace}
\usepackage{tikz}
\usepackage{qtree}
\usepackage{tikz-qtree}
\usepackage{subfigure}
\usepackage{algpseudocode}
\usepackage{algorithm}
\usepackage{multirow}
\usepackage{tcolorbox}
\usepackage{xcolor}
\usepackage{tabularx}
\usepackage{program}
\usepackage[colorlinks=true, allcolors=blue]{hyperref}
\usepackage{listings}
\usepackage{amsmath}
\usepackage{graphicx}

\definecolor{codegreen}{rgb}{0,0.6,0}
\definecolor{codegray}{rgb}{0.5,0.5,0.5}
\definecolor{codepurple}{rgb}{0.58,0,0.82}
\definecolor{backcolour}{rgb}{0.95,0.95,0.92}

\lstdefinestyle{mystyle}{
    backgroundcolor=\color{backcolour},   
    commentstyle=\color{codegreen},
    keywordstyle=\color{magenta},
    numberstyle=\tiny\color{codegray},
    stringstyle=\color{codepurple},
    basicstyle=\ttfamily\footnotesize,
    breakatwhitespace=false,         
    breaklines=true,                 
    captionpos=b,                    
    keepspaces=true,                 
    numbers=left,                    
    numbersep=5pt,                  
    showspaces=false,                
    showstringspaces=false,
    showtabs=false,                  
    tabsize=2
}

\title{CDL: A fast and flexible library for the study of \\ permutation sets with structural restrictions}

\date{}

\author[1]{Bei Zhou}
\author[2]{Klas Markstr{\"o}m} 
\author[1]{S\o ren Riis\thanks{Corresponding author: s.riis@qmul.ac.uk}}

\affil[1]{Queen Mary University of London}
\affil[2]{Ume\aa\ University}

\begin{document}
\maketitle

\begin{abstract}
In this paper, we introduce CDL, a software library designed for the analysis of permutations and linear orders subject to various structural restrictions. Prominent examples of these restrictions include pattern avoidance, a topic of interest in both computer science and combinatorics, and "never conditions" utilized in social choice and voting theory.

CDL offers a range of fundamental functionalities, including identifying the permutations that meet specific restrictions and determining the isomorphism of such sets. To facilitate exploration of large permutation sets or domains, CDL incorporates multiple search strategies and heuristics.
\end{abstract}

\section{Motivation and significance}
\label{sec:motivation}
Permutations, and equivalently linear orders, which avoid various small structures, show up in a wide range of research areas. In the theory of algorithms, they appeared in Knuth's 1968 result \cite{knuth1968art}, which showed that a permutation can be sorted using a stack if and only if the permutation avoids the pattern 231, and later on \cite{simion1985restricted} started an entire research area in combinatorics. In social sciences, sets of linear orders, called domains, appear as the sets of possible rankings of election candidates. \cite{black48} gave an example of a structural restriction that gives a domain where majority voting behaves well, in contrast to Arrow's result \cite{Arrow1951}, which shows that majority voting can misbehave in various ways without such restriction. \cite{sen66} gave the general form for such voting restrictions as never conditions. The global structure and maximum possible size of such domains have been followed up in a large body of work, surveyed in \cite{monjardet2009acyclic,karpov2022structured,puppe2022maximal}.
The large and lively research fields stemming from these early examples have also overlapped, as computational social choice leads to intricate questions in computational complexity \cite{elkind2022}.

The CDL library, originally named Condorcet Domain Library, was first developed to study Condorcet domains, i.e. domains where majority voting leads to a transitive ranking of the candidates. However, since work on pattern-restricted permutations and Condorcet domains require the same underlying functionality, the library was extended to support both research areas and include even more general functions than used.

CDL can be used to find the domain consisting of all permutations that satisfy a prescribed set of constraints easily and look for constraints leading to domains with different properties.  The library can also be used to analyse the structure of a domain, test if domains are isomorphic and visualize domains.  A Python interface is available for ease of use, while the speed-sensitive low-level functions are coded in C\texttt{++}.

\section{Software description}
\label{sec:description}
The Condorcet Domain Library is a comprehensive and versatile tool designed to work with Condorcet domains and forbidden permutations. The library has interfaces for C\texttt{++} and Python programming languages, allowing users seamless library integration into their projects while leveraging its power in their preferred programming environment. The library's modular design also makes customisation easy according to project-specific requirements. The low-level functions of CDL are implemented in C\texttt{++} and follow best practices for efficient and reliable computation. 

The basic objects for the library are \emph{domains}, our common name for a set of permutations satisfying some constraints. The permutations are always on a set $X_n=\{1,2,\ldots,n\}$  where $n \in \{1,2,3,\ldots\}$ and we refer to the elements of $X_n$ as either symbols or alternatives.

The most general form of our constraints is a \emph{scheme}, a collection of $k$-tuples of the alternatives and, for each tuple, a collection of forbidden sets of permutations of those alternatives. The latter is called a \emph{law}.

Pattern-avoiding permutations give the simplest form of a scheme. All $k$-tuples have a law corresponding to a single forbidden permutation of those $k$ alternatives. That law comes from a single forbidden pattern, such as a monotone increasing sequence of length $k$. For Condorcet domains, the $3$-tuple is called a triple and the laws are derived from \emph{never conditions}. A never condition $xNi$ for a triple of alternatives specifies that the $i$th element of the triple cannot appear in position $x$ when the permutation is restricted to those three alternatives. For Condorcet domains, the never conditions typically vary between different triples. A never condition is also commonly called a \emph{never rule} or a \emph{rule}.

\subsection{Software architecture}
The library contains two main classes, which are \texttt{CondorcetDomain} providing the functionalities for Condorcet domain-related calculation, in which the triples and their rule are stored in a \texttt{TRS} object, and \texttt{ForbiddenPermutation} for working with forbidden permutations on $k$-tuples, in which the $k$-tuples and their laws are stored in a \texttt{TLS} object. A \texttt{TRS} or \texttt{TLS} is full if all the $k$-tuples in it are assigned with a rule, and the resulting domain computed from it is called a full domain. If a domain is derived from partial \texttt{TLS} or \texttt{TRS} where at least one $k$-tuple is unassigned, it is called a partial domain \cite{zhou2023new}.  The functionalities offered by these two classes mirror each other, besides a few minor necessary differences in the interfaces. The terminologies we use in this section are consistent with the ones used in \cite{zhou2023new, akello2023condorcet}. We refrain from discussing the mathematical rationales behind the functions but instead lay focus on the implementation aspects. In this section, we show the core ingredients of this library and recommend readers visit our GitHub page for more information\footnote{GitHub. \url{https://github.com/sagebei/cdl}}.

\subsection{Software functionalities}

\subsubsection{Ordering $k$-tuples, rule initialization and assignment}
When constructing the domain defined by a set of constraints, the ordering of the set of $k$-tuples is crucial. A well-chosen ordering can rapidly reduce the size of partial domains during the search. Different orders of these tuples possess distinct properties that can be exploited. To this end, the library inherently supports two orders for general $k$-tuples, namely lexicographic (Lex) order (\texttt{init\_trs\_lex}), co-lexicographic (CoLex) order (\texttt{init\_trs\_colex}), and one additional RZ-order for triples (\texttt{init\_trs}) proposed in \cite{zhou2023new}.

The lexicographic order dictates that $k$-tuple $\{x_1,x_2,\dots,x_k\}$ is before $\{y_1,y_2,\dots,y_k\}$ if $x_1\leq y_1$ and $x_2 \leq y_2$, $\dots$, and $x_k \leq y_k$. The Colexicographic order specifies that $k$-triple $\{x_1,x_2,\dots,x_k\}$ is before $\{y_1,y_2,\dots,y_k\}$ if $x_k\leq y_k$ and $x_2 \leq y_2$, $\dots$, and $x_1 \leq y_1$. The CoLex order has the property that all triples that contain $n+1$ come after the triplets from $\{1,2 \dots n\}$.  The RZ-order for triples is specifies that triple $\{x_1,x_2,x_3\}$ is before $\{y_1,y_2,y_3\}$ if $x_1<y_1$ or $(x_1=y_1$ and $x_3<y_3)$ or $(x_1=y_1$ and $x_2=y_2$ and $x_3<y_3)$. 

The three ways to initialize a list of $k$-tuples leave all unassigned. The library provides two methods to associate a tuple with a rule. The first method uses \texttt{assign\_rule} that assigns a given rule to the specified tuple. The second method, \texttt{assign\_rule\_by\_index} assigns a given rule to the tuple with a specified index where the index is the location of the tuple in the \texttt{TRS}.

CDL also provides the flexibility of defining customized schemes to initializing rules. Initializing rules with a scheme plays an important role in constructing large Condorcet domains, for instance, the alternating scheme \cite{fishburn1997acyclic} and the set-alternating scheme \cite{karpov2023set}. The \texttt{init\_trs\_by\_scheme} initializes rules with a customized function that takes a triple as a parameter and returns the rule assigned according to the scheme. 

For Condorcet domains, we have also implemented a new approach named \texttt{dynamic\_triple\_orde \allowbreak ring} that determines the next triple to be assigned a rule dynamically. The aim is to find the triple that leads to a (partial) CD with the smallest size possible among the unassigned ones. This approach entails keeping a list of the size of partial CDs for all the unassigned triples when assigned with one of the four rules that maximize the size of this partial CD and choosing the triple corresponding to the smallest domain size in that list. 

\subsubsection{Domain construction and size calculation}
Constructing a domain, or calculating its size, from a list of $k$-tuples with a rule assigned is one of the core functionalities provided by this library. Starting with a domain on 2 alternatives $\{\{1, 2\}, \{2, 1\}\}$, the \texttt{domain} function adopts a breadth-first search method that iteratively expands the domain by inserting a new alternative to every position in the permutations and discarding the ones that violate one of the laws. 

Building a domain for the sole purpose of determining its size is impractical when the number of alternatives $n$ is large since the memory usage is proportional to the size of the resulting domain, which on average grows exponentially with the parameter $n$.  To address this issue, we provide the \texttt{size} function, which employs depth-first search to count the number of permutations in a domain. 

The \texttt{size} function utilises an extension function that takes a single permutation and a list of (partially) assigned $k$-tuples and returns the permutations with one added alternative that satisfies all the rules. Inside the \texttt{size} function, a global counter keeping track of the remaining permutations is first set. Given a domain with two initial permutations, the extension function is applied to each permutation in the list and then recursively repeats for each resulting permutation. If a valid permutation on all $n$ alternatives is found, the size counter is increased by 1, and afterwards, the permutation is discarded. When the computation is finished, the counter records the size of the resulting domain. Counting the size this way eliminates most of the memory usage incurred in the \texttt{domain} function and makes it possible to calculate the size for large alternatives in practice. 

Furthermore, a reverse function named \texttt{domain\_to\_trs} is also available, which given a domain, uncovers the rules satisfied by the list of $k$-tuples. Another usage of this function is to reveal the implied rules. Some combinations of the rules in the full \texttt{TRS} could imply a new rule for some triples, i.e. some triples satisfy more never conditions than those appear in the \texttt{TRS}. These new hidden rules are called implied rules. The implied rules in a \texttt{TRS} can be uncovered by first calculating its corresponding domain by the \texttt{domain} function and then constructing a new \texttt{TRS} by \texttt{domain\_to\_trs} function which contains all the rules, both existing and hidden.

\subsubsection{Subset-restriction functions}
A domain on $n$ alternatives can be restricted to a subset of $t$ alternatives. Given a \texttt{TRS} object with $n$ alternatives created by \texttt{init\_trs} function,  we provide the \texttt{subset\_trs\_list} that takes it as input and returns a list of \texttt{TRS} restricted to $t$ alternative subsets, initialized with \texttt{init\_subset \allowbreak (sub\_n=t)}. Prior to using any subset functions, it is mandatory to set the number of subset elements by \texttt{init\_subset(sub\_n=t)}. Built on top of \texttt{subset\_trs\_list}, the \texttt{subset\_states} further computes the state of each of the subset \texttt{TRS}. 

We define \texttt{state} as a numeric representation of a list of rules assigned to triples in order and provide \texttt{trs\_to\_state} function that converts the list of never rules in a \texttt{TRS} object to a list of numbers as well as the \texttt{state\_to\_trs} function that achieves the opposite. For a list of $k$-tuples in lexicographic or co-lexicographic order or customized order, the \texttt{subset\_states\_any\_ordering} function returns a list of subset states where the $k$-tuples are in the RZ-order. 

Computing the subset domains directly from a domain is supported by the \texttt{subset\_domain\_list} function which, given a domain, returns a list of its subset domain on \texttt{sub\_n} alternatives where \texttt{sub\_n} is specified in the \texttt{init\_subset(sub\_n=t)} function. 

The concepts of \emph{ampleness} \cite{puppe2022maximal} and \emph{copiousness} \cite{slinko2019condorcet} have been introduced on the restriction of CDs. To this end, we provide two functions \texttt{is\_domain\_ample} and \texttt{is\_domain\_copious} to facilitate checking the ampleness and copiousness of a domain. \cite{karpov2023local} generalised these two concepts to the concept of local diversity of CDs, the calculation of which is supported by the \texttt{subset\_domain\_list} function.

\subsubsection{Hashing and identifying non-isomorphic domains}
For Condorcet domains and more general domains from social choice and voting theory one often wants to identify isomorphic domains. Here, two domains $D_1$ and $D_2$ are isomorphic if there is a bijection from the set of alternatives of $D_1$ to the set of alternatives of $D_2$  which maps $D_1$ to $D_2$.

Using isomorphism also allows us to define a normal form for a domain $D$. Here we take the normal form to be the lexicographically smallest/minimal domain which is isomorphic to $D$.

This normal form is found by the function (\texttt{isomorphic\_hash}) where two full domains are isomorphic if and only if they have the same hash value. The hash function utilizes the \texttt{inverse\_cd} that transforms a domain $\mathbf{A}$ with a permutation $g$, yielding a domain $\overline{\mathbf{A}}$ that is isomorphic to $\mathbf{A}$. Given a domain $\mathbf{A}$, the \texttt{isomorphic\_domains} function generates all of its isomorphic domains. It applies the \texttt{inverse\_cd} function on every permutation in $\mathbf{A}$ and removes duplicate domains, resulting in a list of the domains that are isomorphic to it. 

Built on top of the \texttt{isomorphic\_domains} function, given a domain, the \texttt{isomorphic\_hash} function gets the list of isomorphic domains, sorts them, and returns the smallest domain as its hash value. 

With the hash function, eliminating the isomorphic domains from a list of domains can be achieved by applying it to each and retaining one domain out of these with the same hash value. This ensures that the list of domains is non-isomorphic to each other. But this is inefficient. We instead implement a faster algorithm in \texttt{non\_isomorphic\_domains}. Here, for every domain $\mathbf{A}$ in the list the function generates the isomorphic domain $\overline{\mathbf{A}}$ for every permutation in $\mathbf{A}$ using \texttt{inverse\_cd} function, and then deletes all domains in the list that are identical to $\overline{\mathbf{A}}$, resulting in a list of non-isomorphic domains.

Additionally, the library provides another function named \texttt{is\_trs\_isomorphic} that given a \texttt{TRS} object, checks if it is lexicographically minimal. Assuming the list of the triples and their laws are given an ordering such that they can be sorted lexicographically, meaning that any (partial) domain can be represented by a list of laws for triples and consequently two domains can be compared by comparing their list of laws lexicographically. Taking a permutation $g$, we can apply it to the alternatives in a triple and its assigned rule to get a new triple and a new rule for it. A triple $\{i,j,k\}$ with a rule $xNp$ is transformed to $\{gi,gj,gk\}$ with rule $gxNp$. Applying this to all triples and their assigned rule, and then listing the new triples according to the predefined order on them, leads to a new list of triples with their new rule which can be compared with the original list. 

The \texttt{is\_trs\_isomorphic} function can be used in searches to remove partial domains that is isomorphic to a domain in another part of the search tree \cite{markstrom2024arrows}.

\subsubsection{Native support for general forbidden permutation domains}
Both sets of pattern-restricted permutations and Condorcet domains are special cases of general forbidden permutation domains. For Condorcet domains, any never condition can be expressed as a pair of forbidden permutations. For example, the condition $3N1$ can be translated into two forbidden permutations $312$ and $321$. 

Generally forbidden permutation domains are implemented via a  class named \texttt{ForbiddenPermutation} that supports core functionalities and interfaces similar to those of the \texttt{CondorcetDomain}. 

Inside \texttt{ForbiddenPermutation}, every $k$-tuple can be assigned with a list of forbidden permutations, which we call laws, differing from that in the \texttt{CondorcetDomain} where every triple is associated with a single never rule. Correspondingly, we have the \texttt{TLS} (denoting TupleLaws) class whose object stores a list of $k$-tuples with their assigned laws.

\section*{Illustrative Examples}
\label{sec:illustrative}
In this section, we provide an overview of some of the core functionalities provided by this library with two examples using \texttt{CondorcetDomain} and \texttt{ForbiddenPermutation} classes, demonstrating a way to put them into practice.  Readers can visit our GitHub page that houses this information for further details, examples, and explanations. 

\subsection{Condorcet domains}
In the code snippet below, the import statement at the first line loads all the classes and functions in CDL into the program. Following this, we define the alternating scheme as a function that takes input as a triple and returns the rule assigned to it. To work with 8 alternative domains, the \texttt{CondorcetDomain} object is initialised with the \texttt{n} parameter set to 8. The \texttt{init\_trs\_by\_scheme} function taking the \texttt{alternating\_scheme} function as the parameter creates a \texttt{TRS} object that stores a list of triple whose rule is assigned as per the alternating scheme.

The following \texttt{domain} and \texttt{size} functions build up the domain and calculate its size in totally different ways as described in the above section. To modify the rule assigned to a triple, \{2, 3, 4\} for instance, we can use the \texttt{assign\_rule} method to change its assigned rule from $2N3$ to $3N1$. \texttt{subset\_states} function finds all the subset states restricted to 6 alternatives, given the \texttt{sub\_n} value being set to 6 in the \texttt{init\_subset} function.  

At the end, we show that given a list of domains, the \texttt{non\_isomorphic\_domains} removes all the isomorphic domains and returns the nonisomorphic ones. 

\begin{lstlisting}[language=Python]
from cdl import *

def alternating_scheme(triple):  # build the alternating scheme 
    i, j, k = triple
    if j % 2 == 0:
        return "2N1"
    else:
        return "2N3"

cd = CondorcetDomain(n=8)  # initialize the Condorcet domain object
# initialize the trs with the predefined alternating scheme 
trs = cd.init_trs_by_scheme(alternating_scheme)  
domain = cd.domain(trs)  # construct the Condorcet domain
# calculate the size of the resulting Condorcet domain (222)
size = cd.size(trs)  
assert len(domain) == size  # True

# change the rule assigned to the triple [2, 3, 4] from "2N3" to "3N1"
trs = cd.assign_rule(trs, [2, 3, 4], "3N1")  
size = cd.size(trs) # the size of the new domain is 210.

cd.init_subset(sub_n=6)
# get a list of 28 subset states in 6 alternatives
substates = cd.subset_states(trs)

# build a list of domains
domains = [cd.domain(cd.init_trs_random()) for _ in range(100)]  
# filter out the isomorphic domains
non_isomorphic_cds = cd.non_isomorphic_domains(domains)
\end{lstlisting}

\subsection{Pattern restricted permutations}
The functionalities and naming conventions of the \texttt{ForbiddenPermutation} class are consistent with those in the \texttt{CondorcetDomain} class, besides the restrictions are a set of laws. In the following code block, we give an example showing how to use the functions in the \texttt{ForbiddenPermutation} to find the size of the domains for 5 to 10 alternatives on $5$-tuples avoiding the forbidden permutation $\{2, 5, 3, 1, 4\}$. 

\begin{lstlisting}[language=Python]
from cdl import ForbiddenPermutation

for n in range(5, 11):
    # initialize the ForbiddenPermutation object for 5-tuples
    fp = ForbiddenPermutation(n, 5) 
    tls = fp.init_tls()
    for tl in tls:
        # assign all the 5-tuples with the law [2, 5, 3, 1, 4]
        tls = fp.assign_laws(tls, tl.tuple, [[2, 5, 3, 1, 4]])
    print(fp.size(tls))
\end{lstlisting}

This code structure closely resembles the one working with CDs. The \texttt{size} and \texttt{domain} functions in the \texttt{ForbiddenPermutation}  class have been used to verify known results for permutations that avoid any of the three inequivalent length 4 patterns (A022558, A061552, and A005802) and the patterns of length 5 presented in \cite{clisby2021classical}.

\section{Impact}
\label{sec:impact}

\cite{karpov2023set} introduced a new class of Condorcet domains constructed via a method called a set-alternating scheme. They used CDL, in particular the \texttt{size} function, to calculate the sizes of the resulting domains. Due to the volume of computation, the parallel functionalities of CDL were utilized on a Linux cluster. 

A new heuristic search algorithm \cite{zhou2023new} was developed using CDL. This algorithm uses a user-defined score function to search for large Condorcet domains and led to the discovery of record-size Condorcet domains for 10 and 11 alternatives, shown in Table \ref{tab:records}. The algorithm also recreated the largest Condorcet domain for n=8 found in \cite{leedham2023largest}, using CDLs functions for parallelized search. Ascribing to the effectiveness of this new algorithm, it was later added to the CDL, allowing users to design customized score functions and integrate the search algorithm into their projects. When this algorithm was added to CDL a depth-first search version of the original algorithm was also developed. This was done in order to address a limitation of the original algorithm where exhaustive searches sometimes consumed too much memory. 

Using the CDL \cite{zhou2023new} also tested the performance, for constructing large Condorcet domains, of a wide range of machine learning algorithms, including deep reinforcement learning algorithms, greedy algorithms, evolutionary algorithms, and local search algorithms. Performance data was presented in a way that makes it easy to evaluate new search algorithms. 

\begin{table}[H]
\centering
\caption{Known maximal size of Condorcet domains}
\begin{tabular}{ccc}
\toprule
\textbf{n} & \textbf{Alternating scheme} & \textbf{New records} \\ 
\midrule
8 & 222 & 224 \\
9 & 488 & \textbf{492} \\
10 & 1069 & \textbf{1082} \\
11 & 2324 & \textbf{2349} \\
 \bottomrule
\end{tabular}
\label{tab:records}
\end{table}

\cite{akello2023condorcet} released a comprehensive list of non-isomorphic Condorcet domains on 7 alternatives. Their computational results were obtained without using CDL, but the  \texttt{isomorphic\_hash} function was used to give an independent verification of the isomorphism reduction for the final data set. 

\cite{karpov2023local} used CDL to test the local diversity of Condorcet domains, and computationally prove several theorems. 

Utilising the functions in the CDL library, \cite{markstrom2024arrows} developed a search algorithm that calculated a complete set of non-isomorphic Arrow’s single-peaked domains on $n \leq 9$ alternatives on a supercomputer, enabling them to investigate the richness of such domains. 

\section{Conclusion}
This paper has introduced CDL as a tool for research  on Condorcet domains and pattern-avoiding permutations. 

The core functionalities of this library have been optimized and are stable. However, research in the areas for which CDL is intended is very active and new concepts, and algorithms, appear regularly. In response, we intend to continuously add new functionalities to the library and keep it relevant for the research front of this thriving research field.

\section*{Acknowledgements}
This work was funded by the China Scholarship Council (CSC). This research utilised Queen Mary's Apocrita HPC facility \cite{king2021apocrita}, supported by QMUL Research-IT. 

\bibliographystyle{plainnat}
\bibliography{references}

\section*{Metadata}

\begin{table}[H]
\begin{tabularx}{\textwidth}{lX}
\hline
\textbf{Item} & \textbf{Detail} \\
\hline
Version & 2.2.5 \\
Repository/Executables Link & \url{https://github.com/sagebei/cdl} \\
License & Apache License Version 2.0  \\
Versioning System & git (Code-specific) \\
Languages/Platforms & C++, python, bash; Software: iOS, Linux, Microsoft Windows, Unix-like \\
Compilation/Installation Requirements & Code: cmake, make; Software: pybind11 (automatically downloaded upon installation) \\
Documentation/User Manual & https://github.com/sagebei/cdl \\
Support Email & bei.zhou@qmul.ac.uk \\
\hline
\end{tabularx}
\caption{Combined Metadata for Code and Software}
\label{tab:combined_metadata}
\end{table}

\end{document}